\documentclass[12pt]{iopart}

\usepackage{graphicx}
\usepackage{color}
\usepackage{cite}
\newcommand{\blue}{\color{black}}

\begin{document}

\title[Influence of domain walls on all-optical toggle switching]{Influence of magnetic domain walls on all-optical magnetic toggle switching in a ferrimagnetic GdFe film}
\author{Rahil Hosseinifar$^1$, Evangelos Golias$^1$\footnote{Present address: MAX IV Laboratory, Lund University,
Fotongatan 2, 22484 Lund, Sweden}, Ivar Kumberg$^1$, Quentin Guillet$^1$, Karl Frischmuth$^1$, Sangeeta Thakur$^1$, Mario Fix$^2$, Manfred Albrecht$^2$, Florian Kronast$^3$, and Wolfgang Kuch$^1$}

\address{$^1$ Freie Universit\"{a}t Berlin, Institut f\"{u}r Experimentalphysik, Arnimallee 14, 14195 Berlin, Germany}
\address{$^2$ Institute of Physics, University of Augsburg, Universit\"atsstra{\ss}e 1, 86135 Augsburg, Germany}
\address{$^3$ Helmholtz-Zentrum Berlin f\"{u}r Materialien und Energie, Albert-Einstein-Stra{\ss}e 15, 12489 Berlin, Germany}

\ead{kuch@physik.fu-berlin.de}
\vspace{10pt}
\begin{indented}
\item[]02 January 2022
\end{indented}

\begin{abstract}
We present a microscopic magnetic domain imaging study of single-shot all-optical magnetic toggle switching of a ferrimagnetic Gd$_{26}$Fe$_{74}$ film with out-of-plane easy axis of magnetization by x-ray magnetic circular dichroism photoelectron emission microscopy.  Individual linearly polarized laser pulses of 800 nm wavelength and 100 fs duration above a certain threshold fluence reverse the sample magnetization, independent of the magnetization direction, the so-called toggle switching.  Local deviations from this deterministic behavior close to magnetic domain walls are studied in detail.  Reasons for nondeterministic toggle switching are related to extrinsic effects, caused by pulse-to-pulse variations of the exciting laser system, and to intrinsic effects related to the magnetic domain structure of the sample.  The latter are, on the one hand, caused by magnetic domain wall elasticity, which leads to a reduction of the domain-wall length at sharp tipped features. These features appear after the optical switching at positions where the line of constant threshold fluence in the Gaussian footprint of the laser pulse comes close to an already-existing domain wall.  On the other hand, we identify the presence of laser-induced domain-wall motion in the toggle-switching event as a further cause for local deviations from purely deterministic toggle switching.
\end{abstract}

\vspace{2pc}

\maketitle

\section{Introduction}
The reversal of magnetization at the fastest possible time scales and microscopic length scales is one of the most important issues for the use of magnetic materials in future data-storage or information-processing devices. One route towards that goal is the control of magnetization by ultrafast laser pulses without resorting to magnetic fields, the so-called all-optical switching (AOS) \cite{1498,Stanciu07,Vahaplar09,1694,Alebrand12,Mangin14,1757,104431,1809,ElHadri16,Takahashi16,Lalieu17,John17,1848,1836}. AOS has been observed in a broad variety of ferri- and ferromagnetic materials, either as single-pulse or multiple-pulse switching.  It can be both, depending on the helicity of the exciting laser pulse or helicity-independent. Particularly interesting is the switching of the magnetization by single laser pulses, which has been achieved in Gd-containing ferrimagnetic materials such as GdFe \cite{1809}, GdFeCo \cite{Stanciu07}, or Gd/Co \cite{Lalieu17} bilayers. Individual laser pulses{\blue, even} with temporal width{\blue s} below 100 fs{\blue ,} can switch the magnetization back and forth, independent of the light polarization \cite{1694}. This all-optical toggle switching is explained by the different speeds of demagnetization of Gd and the $3d$ metals in the ferrimagnetic alloy or in bilayers upon exposure to the laser pulse, together with the conservation of angular momentum during the laser-induced demagnetization \cite{Radu11,Mentink12,Wienholdt13}. Due to the helicity-dependent absorption of the laser fluence in the sample as a consequence of magnetic circular dichroism and a sharp intensity threshold for toggle switching, this can also lead to helicity-dependent all-optical single-pulse switching \cite{1694,Khorsand12}.

Important for applications of all-optical toggle switching is the reproducibility of the switching event on microscopic length scales. The presence of magnetic domain walls has been found to limit the repeatability of the switching event, leading to nondeterministic switching in the vicinity of the domain walls \cite{1809}. This has been attributed to thermally activated domain-wall motion. In another study, it has been demonstrated that ultrafast laser pulses can indeed move magnetic domain walls, which however could not simply be explained by thermal activation due to the transient heating by the laser pulse alone, but had to be attributed to a two-step process based on laser-pulse-induced depinning of domain walls and successive thermal domain wall motion after the laser pulse \cite{1772}. In a GdFe-containing sample, in addition, even unidirectional domain-wall motion in the temperature gradient created in the Gaussian footprint of individual ultrashort focused laser pulses has been reported \cite{1825}.

In this paper, we investigate nondeterministic all-optical toggle switching of a Gd$_{26}$Fe$_{74}$ film with out-of-plane easy axis of magnetization by magnetic imaging using photoemission electron microscopy (PEEM) with X-ray magnetic circular dichroism (XMCD) as magnetic contrast mechanism.  We focus single ultrashort infrared laser pulses of 100 fs temporal width to the vicinity of magnetic domain walls and study the local deviations from a perfect, deterministic toggle switching and their relation to the domain-wall position {\blue for temperatures above and below the ferrimagnetic magnetization compensation temperature}.  Besides pulse-to-pulse variations in laser fluence, we identify two main mechanisms which can lead to non-deterministic switching on the nanoscale: (i) domain coarsening, i.e., the reduction of domain-wall energy by avoiding sharp features created in a switching event, and (ii) laser-induced domain-wall motion.  {\blue From our results, we cannot deduce an effect of the base temperature.}  Both {\blue effects} can become important for optical writing of magnetic information. 

\section{Results and Discussion}

\begin{figure}
	\centering
  \includegraphics[width=15cm,keepaspectratio]{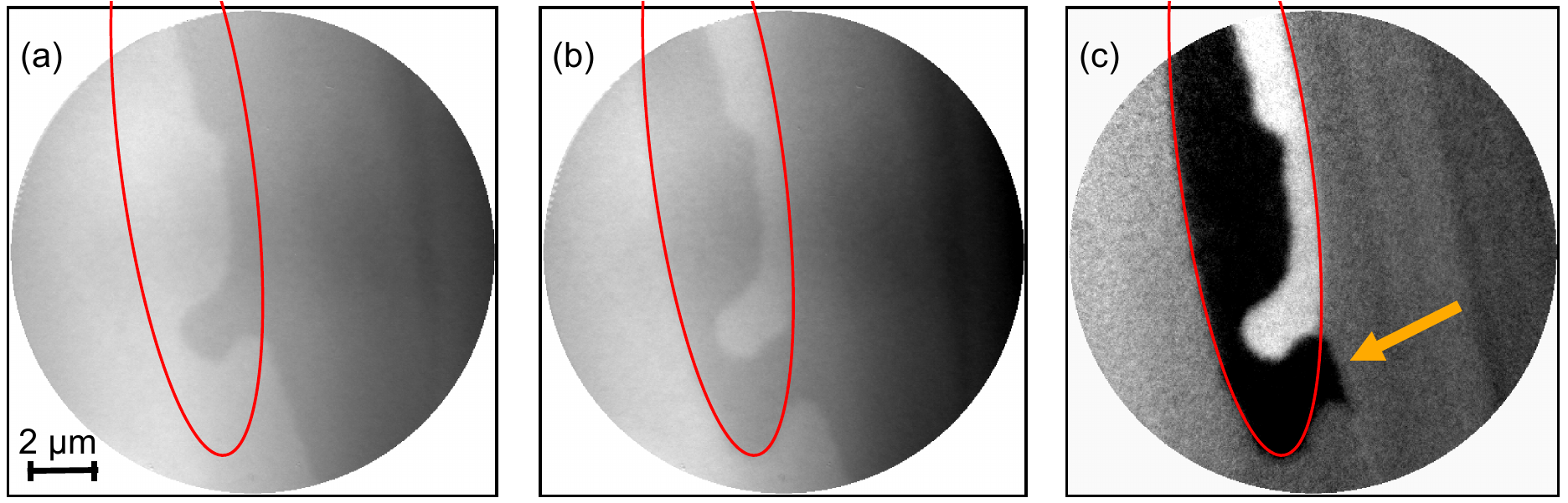}
\caption{Static XMCD-PEEM images acquired at 50 K before (a) and after (b) a single laser pulse of 6.8 mJ/cm$^2$ fluence in the center. Light and dark gray contrast corresponds to opposite magnetization directions.  (c): Difference image (image (b) minus image(a)).  The red ellipse marks a line of constant fluence equaling the threshold fluence for all-optical toggle switching. The yellow arrow indicates a position at which a deviation from the deterministic toggle switching is observed.}
	\label{Fig1}
\end{figure}

Comparison of microscopic magnetic domain patterns, recorded in static XMCD-PEEM imaging before and after excitation of the sample by a single laser pulse in absence of any magnetic field reveals the lateral distribution of optically switched areas.  Figure \ref{Fig1} presents an example.  Panels (a) and (b) show the domain structure before and after one single laser pulse of 6.8 mJ/cm$^2$ incident fluence in the center for a sample temperature of 50 K{\blue , respectively, below the magnetization compensation temperature of the sample of about 120 K}. The red ellipse indicates a line of constant fluence of 6.1 mJ/cm$^2$ of the laser on the sample.  Light and dark grey contrasts indicate the two opposite magnetization directions.  Figure \ref{Fig1} (c) shows the pixel-by-pixel difference of the two images.  Here, black and white contrast corresponds to areas on the sample where the magnetization has switched, either from bright to dark or vice versa, while intermediate grey indicates unchanged magnetization.  It can be clearly seen that magnetic toggle switching is present everywhere inside the marked ellipse, i.e., at fluences above 6.1 mJ/cm$^2$ up to at least the maximum fluence of 6.8 mJ/cm$^2$.  Outside the ellipse, i.e., at fluences below 6.1 mJ/cm$^2$, no switching occurs except in the area marked by an arrow in Fig.\ \ref{Fig1} (c).  We can thus assign the value of 6.1 mJ/cm$^2$ to the threshold for toggle switching in this sample at 50 K.  

The local microscopic deviation from a deterministic toggle switching {\blue defined by a uniform threshold fluence, as} indicated by the arrow in Fig.\ \ref{Fig1} (c){\blue ,} can be explained by reduction of domain-wall energy:  A completely deterministic switching only inside the red ellipse would have left a sharp tip of the bright domain outside the ellipse at this position [Fig.\ \ref{Fig1} (a)].  Domain coarsening by rounding off this sharp tip can save domain wall energy with a relatively small motion of domain walls.  In other words, the elastic pressure on the domain wall, which is proportional to the derivative of the domain-wall length to its position, is very high at such sharp features and may lead to a domain-wall motion, possibly facilitated by the transiently higher temperature shortly after the laser pulse, or by creep motion of the domain wall at longer times after the pulse.  Such a creep motion of sharp domain ends has been studied in detail by Cao {\it et al.} in Co/Gd stacks sandwiched by Pt or Ta \cite{1857}.  

\begin{figure}
	\centering
\includegraphics[width=15cm,keepaspectratio]{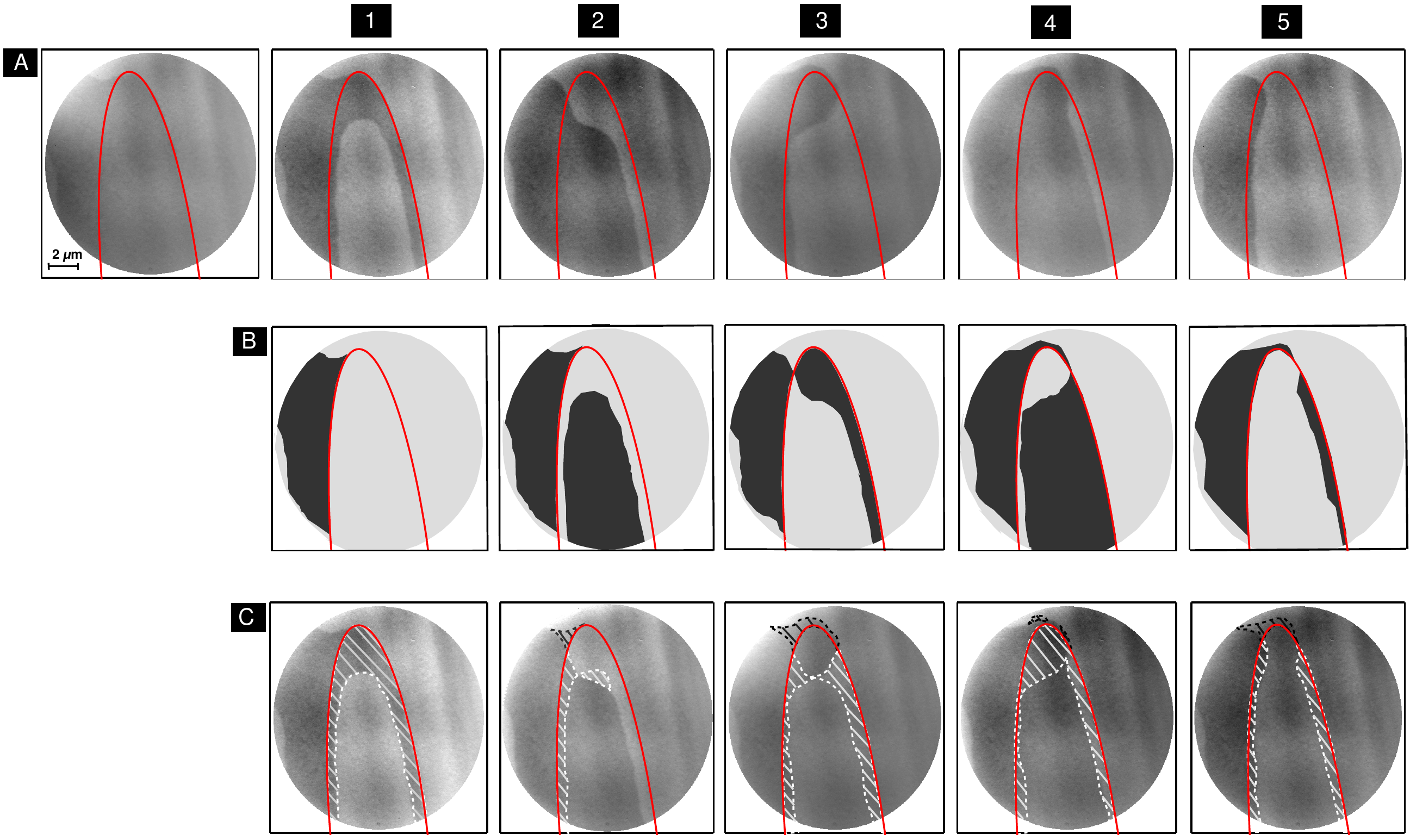}
\caption{Static XMCD-PEEM images acquired at room temperature, each image after one single laser pulse of 800 nm wavelength and fluence 4.9 mJ/cm$^2$ in the center.  Dark and light gray contrast corresponds to opposite magnetization directions. Row A presents a sequence of images with one single laser pulse in between each image.  The  red ellipse marks a line of constant fluence of 3.5 mJ/cm$^2$. Row B presents synthetic images indicating what would be expected if the toggle switching was deterministic inside the red ellipse.  Row C shows the difference between the actual experiment and such a deterministic toggle switching with threshold of 3.5 mJ/cm$^2$.  The white hatched areas indicate regions inside the ellipse which should have switched but did not switch, while the black hatched areas outside the ellipse mark regions of the sample which should not have switched but did switch. The numbers on top of each column count the laser pulses on the sample.}
\label{Fig2}
\end{figure}

To further investigate the effect of magnetic domain patterns on the toggle switching, different parts of the sample were excited by a sequence of individual laser pulses. Figure \ref{Fig2} presents an example of a sequence of PEEM images which were taken before and after successive single laser pulses of the same fluence, now with the sample at room temperature.  The first row, labeled ``A'', shows the PEEM images taken between each single laser pulse.  The red ellipses in each panel show a line of constant fluence of 3.5 mJ/cm$^2$ on the sample, dark and light gray contrast corresponds to opposite magnetization directions.  The second row, labeled ``B'', contains synthetic images sketching the magnetic domain images one would expect if everywhere inside the ellipse the magnetization direction had reversed, while it had remained constant outside.  This would correspond to deterministic toggle switching with a threshold fluence marked by the ellipse.  {\blue It was obtained by first discriminating the dark and bright contrast of the image from the previous column of row A into black and white and subsequently reversing this black-and-white contrast inside the red ellipse.}
The third row, labeled ``C'', repeats the images from row ``A'' while indicating the deviation between the actually observed domain pattern and the one sketched in row ``B''.  White hatched areas indicate those parts of the sample that should have switched but did not and the black hatched areas those that should not have switched but did.

From the images shown in Fig.\ \ref{Fig2}, certain deviations from a purely deterministic toggle switching inside an elliptical area defined by a uniform threshold fluence are evident.   We can distinguish extrinsic and intrinsic effects.  We define extrinsic effects as those being caused by everything outside the sample.  Here, these are most likely shot-to-shot variations of the laser intensity, which could be caused by instabilities in the synchronization of the pulse picker {\blue (the shot-to-shot variation of the laser system itself is typically only around 1\%)}.  The result of shot 1 in Fig.\ \ref{Fig2} on the domain pattern of the sample can be attributed to such an external effect.  The switched area on the sample has a purely elliptic shape and can be explained by a reduced laser fluence in this shot.  Note that the choice of the value of 3.5 mJ/cm$^2$ to place the threshold lines in Fig.\ \ref{Fig2} is somewhat arbitrary considering a variation of the pulse fluence and the relatively small number of five pulses.  {\blue The center position of the laser spot on the sample was determined before and after each experiment as described in the experimental section and did not change within the experimental error of about 0.5 $\mu$m.}

Intrinsic effects concern properties of the sample itself.  Such effects are seen in Fig.\ \ref{Fig2} around the top of the ellipse after laser pulses 2 and 3.  After laser pulse 2, due to the domain formation caused by pulse 1, a deterministic switching inside the ellipse would leave a bright domain with a sharp tip inside the ellipse as well as a dark domain with a sharp tip outside the ellipse (panel B2). The actual situation observed after pulse 2, as shown in row A, is a smooth connection between the dark domains inside and outside the ellipse, reducing the overall length of the domain wall between bright and dark domains and thus the domain wall energy, as discussed already before in connection with Fig.\ \ref{Fig1}.
Such situations occur if already-existing domain walls are close to or cross the constant fluence line related to the threshold fluence for toggle switching.

The same effect in the converse sense occurs at around the same position after the third pulse.  Here the sharp tips expected at the end of the bright domains inside and outside the ellipse at the top left end are rounded off, leaving a rounded bulge of the dark domain (panel A3).  This again reduces the domain-wall length and energy.  In addition, the thin stripe of dark domain expected along the right edge of the ellipse is not present.  The latter might be also due to an extrinsic effect, but since there is no corresponding effect seen on the left side of the laser pulse, that would mean that the position of the laser spot on the sample has changed, which is less likely than a small shot-to-shot variation of the fluence.  It can be also attributed to a shrinking of a thin pointed domain, possibly starting from some position on the bottom right outside the field of view. 

\begin{figure}
	\centering
  \includegraphics[width=15cm,keepaspectratio]{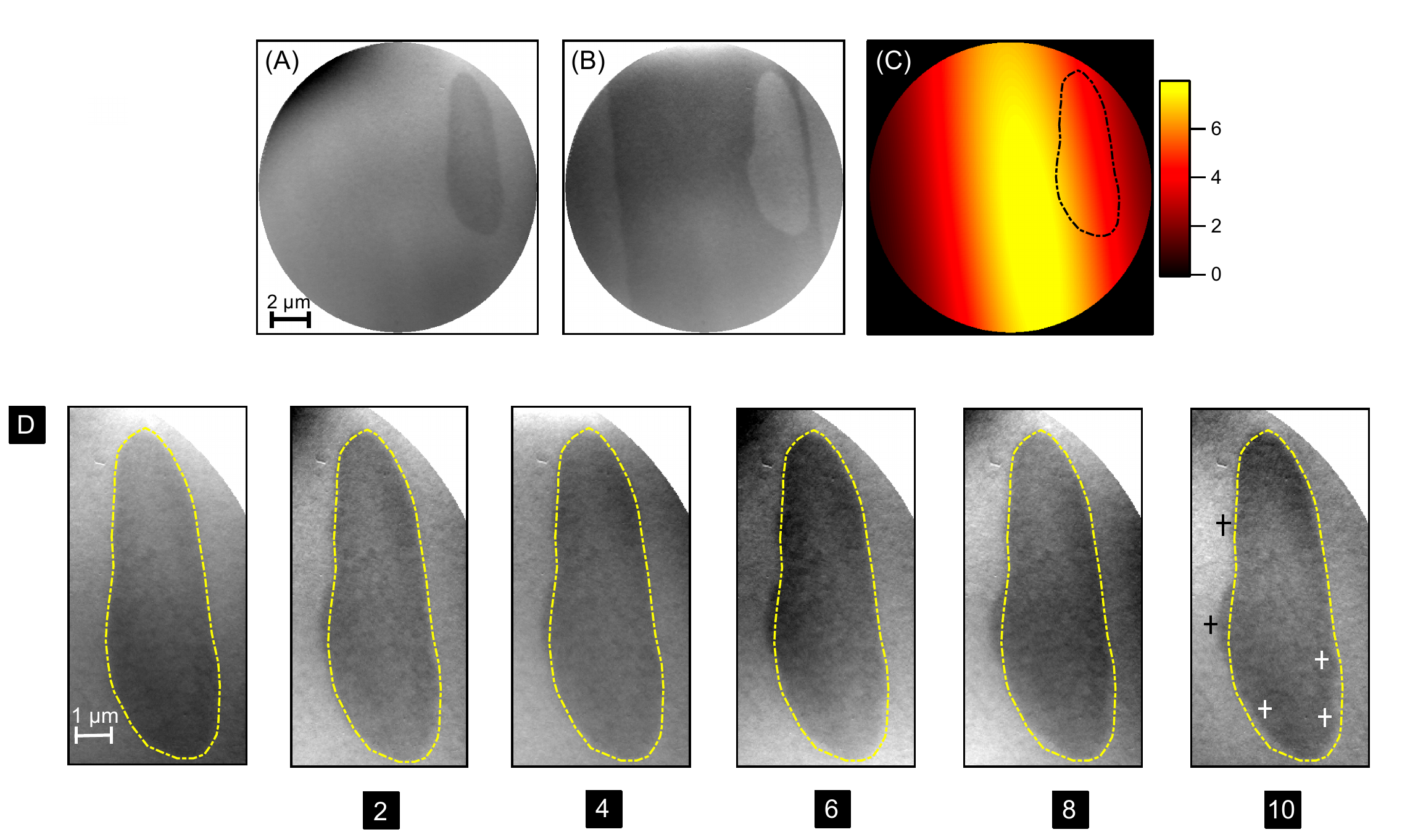}
	\caption{PEEM images before (A) and after (B) a single laser pulse of 7.9 mJ/cm$^2$ fluence in the center at 200 K sample temperature.  (C) shows the laser fluence as a two-dimensional Gaussian profile. Row D shows the configuration of the single domain after different numbers of laser pulses as given at the bottom of each image. The yellow dashed line indicates the initial position of a magnetic domain.}
	\label{Fig3}
\end{figure}

After the fourth pulse, the entire upper tip of the ellipse remained with dark contrast.  This also saves domain-wall energy, since now the existing dark domain outside the top end of the ellipse connects to the dark domain resulting from the toggle switching in the center part of the ellipse.  What is remarkable in panel A4 is the unswitched region of the sample at the right edge of the ellipse.  If it were merely an extrinsic effect, it should follow a smooth line defined by the elliptical line of constant fluence in the footprint of the laser spot.  Instead, there is some excursion towards the inside of the ellipse more towards the top, which cannot be explained by a different pulse position nor by a reduction of domain-wall energy.  We attribute this to laser-induced domain-wall motion and will come back to this effect when discussing Fig.\ \ref{Fig3}.  

Finally, after the fifth pulse, the situation around the top of the ellipse is similar to the pulses before:  The dark domain that should remain outside the ellipse shrinks to a rounded shape on the left, reversing in some area outside the ellipse the magnetization from dark to bright.  At the right edge of the ellipse, a uniform bright magnetization is observed and any dark domain that could have possibly been generated at the position where the bright domain at the right edge in panel A4 reaches the most into the ellipse has either not switched or disappeared, similar to the other sharp domain features.  

These domain-wall motion events caused by domain-wall elasticity are observed independent of the temperature, at 50 K (Fig.\ \ref{Fig1}){\blue , 200 K (not shown), and at} room temperature (Fig.\ \ref{Fig2}){\blue , i.e., well below and above the magnetization and angular momentum compensation temperatures}.  Their kinetics, however, may well depend on temperature.

We mentioned already that the boundaries between switched and unswitched regions of the sample are expected to be smooth and follow a line of constant fluence of the laser footprint for entirely deterministic switching.  However, at some places the domain walls exhibit excursions from such a smooth shape, as for example in panel A3 of Fig.\ \ref{Fig2} at the left edge or in panel A4 at the right edge of the ellipse.  We discuss this now with Fig.\ \ref{Fig3}.  It presents XMCD-PEEM images of a magnetic domain that has been positioned completely within the threshold line for toggle switching.  Panel (A) shows the starting configuration, panel (B) the result after three laser pulses with incident fluence of 7.9 mJ/cm$^2$ in the center.  Here the sample is at a temperature of 200 K.
The odd number of pulses switches the domain from dark to bright and the surrounding, within the threshold line for toggle switching, from bright to dark.  The threshold for toggle switching is recognized from the two straight domain walls following the two long sides of the elliptic shape of the laser spot on the sample.  The center of the laser pulse is about in the center of the image.  Panel (C) shows the local incident fluence in the laser spot on the sample and outlines the boundary of the domain under consideration, which is located in the fluence gradient of the laser spot.  

The bottom row of Fig.\ \ref{Fig3} [panel (D)] shows enlarged images of the same domain as in panels (A) and (B).  Between each image, two laser pulses  were applied with a temporal separation of the order of 1 s except for pulses 3 and 4, which had a separation of about 1 min.  
The number of laser pulses applied since the beginning of the series is indicated at the bottom of each image.  The shape of the initial domain is marked by a dotted yellow line in all images.  For a completely deterministic switching, all the images should be identical.  However, clear differences on the scale of below 0.5 $\mu$m are observed.  Already after two pulses, a hump in the left domain wall becomes obvious, marked by {\blue the lower one of the two} small black ``+'' sign{\blue s} in the last image.  It cannot be explained by domain coarsening, since the overall domain-wall length is not reduced but rather enhanced by that excursion.  
A few more, smaller, changes are observed also at the top left of the domain{\blue , marked by the upper black ``+'' sign in the last image, and at the lower right end of the domain, marked by two white ``+'' symbols, where the domain wall develops two inward excursions}.  In all these places the domain walls have moved to the left after the two laser pulses.  All in all, the left boundary of the domain is clearly less straight after the first two pulses compared to the initial situation.

Continuing the series of laser pulses, these excursions get more pronounced.  In addition, starting from the sixth pulse, a domain-wall movement towards the inside of the domain is observed at the lower left end of the domain, marked by {\blue the leftmost} white ``+'' symbol in the last image.  Except for {\blue this} lower left end of the domain, the movement of the domain walls is towards the side of the higher fluence in the fluence gradient on the sample.  The movement towards the right at the lower left end of the domain might as well be driven by the domain wall energy, since there the observed domain-wall motion reduces the overall domain wall length.  

We interpret this domain-wall motion in accordance with our previous publications as laser-induced depinning and successive thermal domain-wall motion in the locally heated region of the sample \cite{1772,1825}.  The places at which domain-wall motion is observed could comprise of a locally shallower domain wall pinning potential than at other places along the domain wall {\blue and probably reflect the microstructure of the sample}.  The direction of the subsequent domain-wall motion, which requires a time of the order of nanoseconds, is affected by the change in domain-wall energy and possibly also a directional pressure related to the direction of the temperature gradient within the footprint of the laser spot on the sample \cite{1825}.   Although the observed domain-wall motion is predominantly in the direction of higher laser fluence, the data presented here do not allow to prove a directional motion of domain walls, as has been found previously in a Co/Gd$_{25}$Fe$_{75}$ bilayer \cite{1825}.  Opposite to Ref.\ \cite{1825}, here the domain-wall motion, if directed, would be into the direction of higher fluence, not into the direction of lower fluence.  Domain-wall motion towards the higher temperature in a temperature gradient can be explained by entropy \cite{1744,1794} or conservation of angular momentum during the transmission of magnons driven by the temperature gradient through the domain wall \cite{1743}.  The laser-induced domain-wall motion to the colder side in the fluence gradient of the laser spot on the sample as observed in Ref.\ \cite{1825} had been tentatively explained by the conservation of angular momentum upon reflection of magnons at the domain wall.  Whether magnons are predominantly reflected or transmitted at a domain wall depends on the relation between magnon wavelength and domain-wall width.  If the magnon wavelength is large compared to the domain-wall width, magnons are predominantly reflected at the domain wall \cite{Wang2012}.  An alternative explanation of the directed laser-induced domain-wall motion is based on the angular momentum compensation point in ferrimagnets, below which the transfer of angular momentum drives the domain wall in a uniaxial ferrimagnet towards the colder end of the sample above the Walker breakdown \cite{1827}.   In the example of Fig.\ \ref{Fig3}, the temperature of 200 K could {\blue in principle still} be below the angular momentum compensation temperature,
which is higher than the magnetization compensation temperature {\blue (in GdFeCo alloys by even up to about 100 K} \cite{Kim17}).  {\blue However, in Fig.\ \ref{Fig2}, where in panel A4 also an example of such laser-induced domain-wall motion towards the direction of higher laser fluence is observed, the temperature of 300 K should be clearly above both compensation temperatures.}  The difference between the direction of domain-wall motion predominantly observed here {\blue i.e., towards the direction of higher laser fluence,} and in Ref.\ \cite{1825} could then be due to differences in the material parameters of the two samples, such as the presence or absence of the additional Co layer or the different GdFe composition, leading to  different domain-wall widths or, in the frame of the model of Ref.\ \cite{1827}, {\blue to different temperatures relative to the angular momentum compensation temperature}, a different Walker breakdown, or less uniaxial anisotropy.  

\section{Conclusion}

XMCD-PEEM imaging of the deviation from deterministic all-optical toggle switching of a Gd$_{26}$Fe$_{74}$ film by single ultrashort laser pulses showed that the factors that cause this deviation can be distinguished as intrinsic and extrinsic effects.  Variations in the fluence from pulse to pulse as an extrinsic effect is not fully avoidable and can lead to nondeterministic toggle switching close to the fluence threshold line in the footprint of the laser pulse. 
As intrinsic effects we can identify two different mechanisms:
The first one is related to the shape of the magnetic domains as they are left directly after the all-optical switching and the force on the domain wall resulting from the balance of domain-wall and magnetostatic energy.  Sharp domain features that would occur where domain walls are close to the boundary of the optically switched region in the laser spot are avoided by domain-wall motion events. This is observed independent of the sample temperature, below and above the magnetic compensation temperature. 
The second intrinsic effect is laser-induced domain wall motion.  This is clearly identified by local deviations in the position of existing domain walls after toggle-switching events.  
The direction of the laser-induced domain-wall motion is predominantly towards the higher laser fluence, but the data does not allow to prove a unidirectional motion as it has been observed previously in a different sample \cite{1825}.  
Our results show that laser-induced domain-wall motion can occur together with all-optical switching and can contribute to local non-deterministic behavior in the toggle switching.  

\section{Experimental}

The sample was a Gd$_{26}$Fe$_{74}$ rare-earth--transition-metal (RE--TM) alloy film with 15 nm thickness, deposited at room temperature using magnetron sputtering (base pressure $< 10^{-8}$ mbar) from elemental targets. The Ar sputter pressure was kept constant at $3.5 \times 10^{-3}$ mbar during the deposition process. The film was prepared with 5 nm Pt as a seed layer on a Si(100) substrate with a 100 nm thick thermally oxidized SiO$_x$ layer and covered with a 3 nm Al capping layer to prevent oxidation. 
Composition and layer thicknesses were determined by calibrating the sputter rates with a quartz balance before the depositions.
Thus, the sample structure was as follows: Al (3 nm)/Gd$_{26}$Fe$_{74}$ (15 nm)/Pt (5 nm)/SiO$_x$ (100 nm)/Si(100).  
{\blue Magnetic properties were characterized by s}uperconducting quantum interference device vibrating sample magnetometry{\blue , which} confirmed an out-of-plane easy axis of magnetization with rectangular hysteresis loops and a coercivity of about 5 mT at room temperature, see Supporting Information File 1 for temperature-dependent magnetization loops.  

Magnetic domains were resolved using the photoemission electron microscope (PEEM) at the UE49-PGM SPEEM beamline of BESSY II.  The acceleration potential between the sample and the first objective lens of the PEEM was set to 15 keV.  X-ray magnetic circular dichroism (XMCD) at the Gd $M_{5}$ absorption edge at 1182.6 eV photon energy was used as magnetic contrast mechanism.  A small electromagnet mounted inside the sample holder allowed to apply a magnetic field to the sample for demagnetizing or creating domains.  
{\blue Before the start of each series of laser exposures, the electromagnet was used to saturate the sample.  Magnetic domain walls have been created by subsequently applying magnetic fields of reversed polarity.}
All images presented here were recorded with positive helicity of circularly polarized x rays.  In those images, light (dark) gray contrast corresponds to magnetic domains having a positive (negative) projection of their magnetization direction on the x-ray incidence.   
Experiments were performed at a sample temperature of 50 K, which is below the magnetic compensation temperature (${T}_\mathrm{M}$)
of the ferrimagnet, as well as at 200 K and at room temperature, both being above $ T_\mathrm{M}$. 

The sample was excited by laser pulses from a Femtolasers Scientific XL Ti:sapphire oscillator set to about 100 fs temporal width.  A Femtolasers Pulsfinder was used to deliver single laser pulses to the sample.  The linearly $p$-polarized laser pulses with a central wavelength of 800 nm were focused on the sample at a grazing incidence of $16^{\circ}$ to a spot size of about 
$12\times60\ \mu$m$^2$ (at $1/e$ of the maximum intensity) 
by a lens inside the vacuum chamber.  
The combination of a polarizer and a half-wave liquid crystal plate constitutes a variable attenuator for tuning the fluence, which in the center of the spot ranged from 0 to 25 mJ/cm$^2$ incident fluence.  The numbers given for the fluence contain a 15 \% systematic error due to the uncertainty in the determination of the spot size on the sample and the attenuation of the laser light by the optical elements on the way to and into the vacuum chamber.  The position of the laser spot on the sample was determined with an accuracy of 0.2 $\mu$m along the short axis and about 1 $\mu$m along the long axis of the elliptic footprint of the laser beam on the sample before and after acquisition of each series of magnetic domain images at a strongly reduced fluence and 1.25 MHz repetition rate, recording PEEM images of the intensity from three-photon photoemission processes at hot spots of the sample surface and fitting the resulting intensity to a two-dimensional elliptic Gaussian.

\ack
We thank the Helmholtz-Zentrum Berlin for the allocation of synchrotron radiation beamtime.

This work was supported by the Deutsche Forschungsgemeinschaft via the CRC/TRR 227 ``Ultrafast Spin Dynamics", project A07. 

\section*{References}
\bibliography{GdFe_v8}

\end{document}